\documentclass[conference]{IEEEtran}
\usepackage[utf8]{inputenc}

\usepackage{xspace}
\usepackage{url}
\usepackage{graphicx}
\usepackage{listings,multicol}
\usepackage[export]{adjustbox}
\usepackage{amsmath}
\usepackage{amssymb}
\usepackage{textcomp}
\usepackage{breakurl} 
\usepackage{algorithm}
\usepackage{tikz}
\usepackage[noend]{algpseudocode}
\usepackage{balance}

\newcommand{\ie}{\emph{i.e.,}\xspace}
\newcommand{\eg}{\emph{e.g.,}\xspace}

\newcommand{\toolname}{\textsc{Sorrel}\xspace}

\title{\toolname: an IDE Plugin for Managing Licenses \\ and Detecting License Incompatibilities}

\author{
 \IEEEauthorblockN{Dmitry Pogrebnoy,\IEEEauthorrefmark{1}\IEEEauthorrefmark{2} Ivan Kuznetsov,\IEEEauthorrefmark{3} Yaroslav Golubev,\IEEEauthorrefmark{4} Vladislav Tankov,\IEEEauthorrefmark{6}\IEEEauthorrefmark{1}\IEEEauthorrefmark{4} Timofey Bryksin\IEEEauthorrefmark{4}\IEEEauthorrefmark{2}}
    \IEEEauthorblockA{\IEEEauthorrefmark{1}\textit{JetBrains}, \IEEEauthorrefmark{4}\textit{JetBrains Research},   \IEEEauthorrefmark{2}\textit{Saint Petersburg State University}, \\
    \IEEEauthorrefmark{3}\textit{Saint Petersburg Polytechnic University}, \IEEEauthorrefmark{6}\textit{Higher School of Economics}}
    \IEEEauthorblockA{\{dmitry.pogrebnoy, yaroslav.golubev, vladislav.tankov, timofey.bryksin\}@jetbrains.com, kuznetsov3.ia@edu.spbstu.ru}
}

\begin{document}

\maketitle

\begin{abstract}
    Software development is a complex process that includes many different tasks besides just writing code. One of the aspects of software engineering is selecting and managing licenses for the given project. 
    In this paper, we present \toolname---a plugin for managing licenses and detecting potential incompatibilities for IntelliJ IDEA, a popular Java IDE. The plugin scans the project in search of information about the project license and the licenses of its libraries. If the project does not yet have a license, the plugin provides the developer with recommendations for choosing the most suitable open license, and if there is a license, it informs the programmer about potential licensing violations. The tool makes it easier for developers to choose a proper license for a project and avoid most of the licensing errors---all inside the familiar IDE editor.

    The plugin and its source code are available online on GitHub: \url{https://github.com/JetBrains-Research/sorrel}. A demonstration video can be found at \url{https://youtu.be/doUeAwPjcPE}.
\end{abstract}

\section{Introduction}\label{sec:introduction}

With the growing amount of projects using Free Open-Source Software (FOSS)~\cite{ProjectsUsingFOSS, WiredOSSWon}, the problem of license violations in these projects is becoming more relevant. Research shows that developers of open-source projects can commit licensing violations because they are often limited in time and resources and reuse a lot of code~\cite{haefliger2006knowledge}, and also because the system of open-source licenses is very complex, with developers not fully understanding the terms and limitations of specific licenses~\cite{almeida2017software}. Indeed, there are currently more than 400 open-source licenses with different limitations, their extensive list can be found at the Software Package Data Exchange (SPDX) website~\cite{SPDX}. 

Several studies about license violations in open-source projects have been conducted in recent years~\cite{JetBrains_Arcticle, 10.1109/ICPC.2010.48, Romansky2018SourcerersAA, 10.1109/SEW.2012.24}. These studies demonstrated that potential license violations are present in open-source projects on various levels. Our previous such study~\cite{JetBrains_Arcticle} demonstrated that this problem concerns the most popular licenses like permissive Apache-2.0 and restrictive GPL-3.0. In the most extreme cases, incompatibilities between licenses may lead to problems, like the famous  \textit{Google vs Oracle} legal case~\cite{Google_vs_Oracle}. While large companies can consult with legal teams about licensing, small teams and individual developers have to manage licenses by themselves.

Certain solutions exist for detecting and managing licenses inside projects, for example, \textsc{scancode}~\cite{scancode_toolkit}, \textsc{askalono}~\cite{askalono}, \textsc{licensee}~\cite{licensee}, etc. However, all of these tools exist outside of the modern Integrated Development Environments (IDEs) where the actual software development takes place. This means that developers need to go out of their way to check their licenses, which makes the feedback loop longer and complicates license management. Another problem is the analysis of the results collected from such tools: developers have to spend a lot of time understanding the licenses of various libraries and resolving compatibility issues. To overcome all these limitations, we developed a plugin for managing licenses inside IntelliJ IDEA\footnote{IntelliJ IDEA, a popular IDE for Java: \url{https://www.jetbrains.com/idea/}} called \toolname.

In \toolname, we implemented features that help software developers solve problems that commonly arise during the development cycle. The plugin can extract licenses from modules and libraries used in the project and visualize them inside the IDE in a convenient way. The plugin supports 16 most popular licenses that occur in Java projects based on the recent research~\cite{JetBrains_Arcticle}. The gathering of licenses from the modules of the projects is carried out using two detectors: a machine learning classifier and a detector based on the similarity coefficient of texts. The gathering of library licenses is carried out using the features of the IntelliJ Plaform. 

Knowing the licenses of the libraries in a project, \toolname can carry out license compatibility checks. The information about all potential violations can be conveniently found in the side panel of the editor. In the case when the project has no license, the plugin can suggest the license that will adhere to all the limitations of the licenses in the used libraries. Other key features of the plugin include a convenient way of creating licenses and switching between them, as well as highlighting differences between the given license and its standard text.

\toolname is written in Kotlin and is available on GitHub: \url{https://github.com/JetBrains-Research/sorrel}. We believe that it can serve as a handy support for individual developers and small teams in managing licenses in open-source projects. 
\section{Related Work}\label{sec:background}

The task of detecting open-source licenses in projects is solved by a wide array of existing tools. Some of them use rule-based approaches that consist of searching for specific word patterns, others use heuristics and statistical methods.

\textsc{Ninka}~\cite{10.1145/1858996.1859088} was one of the first popular tools for detecting licenses. The tool allows the developer to not only search for the license of a project, but also to search for licenses inside the headers of source code files too. The tool uses a sentence-matching approach written in Perl. Because of this, \textsc{Ninka}'s operation time is rather long.

\textsc{Scancode}~\cite{scancode_toolkit} is a popular tool that detects all SPDX licenses based on heuristics with a rule-based approach. The tool can search for licenses not only in LICENSE-like files, but also in READMEs and other files. It also supports convenient visualization of the detection results. \textsc{Scancode} is written in Python and recently gained a lot of popularity.

\textsc{askalono}~\cite{askalono} is a tool based on statistical methods. The tool compares the given input text to license texts it supports using bigrams to see how similar they are. It then scores each match using the Sørensen–Dice coefficient~\cite{sorenson1948method, DiceCoefficient} and searches for the highest result. One of the core features of \textsc{askalono} is its high speed. 

\textsc{licensee}~\cite{licensee} is a tool written in Ruby that gained a lot of traction because GitHub uses it to detect licenses in GitHub repositories. The tool searches for licenses inside LICENSE files. Firstly, it tries to find an exact match with known licenses, then, if this approach fails, it uses the Sørensen–Dice coefficient, similar to \textsc{askalono}. 

All of the mentioned tools have certain advantages and disadvantages, including differences in operation time, the support of various licenses and formats, as well as the visualization of the results. However, all of them share certain limitations for developers. Firstly, these tools do not natively support detecting license violations between different licenses, as well as suggesting a project license based on the licenses of libraries, both of which are real use-cases for the end-users. Secondly, all these tools are standalone programs, which makes them applicable in a Continuous Integration environment, but difficult to integrate into an IDE workflow. To overcome these limitations, we developed \toolname.

\section{\toolname internals}\label{sec:internals}

\toolname is a tool for working with the licenses of Java projects inside IntelliJ IDEA. 
\toolname is written in Kotlin and currently supports 16 of the most popular Java licenses. The plugin can:

\begin{enumerate}
    \item Detect and recognize the licenses of project's modules.
    \item Detect and recognize the licenses of project's libraries.
    \item Detect incompatibilities between licenses in the project.
    \item Suggest a project license that will be compatible with all of the project's libraries.
    \item Visualize all the licensing information inside the IDE and provide convenient features for managing licenses.
\end{enumerate}

Let us now describe this functionality in greater detail.

\subsection{Detecting Licenses from Text}
\label{sec:detecting_licenses}

An important task in determining licenses is being able to recognize a license given its text. \toolname uses two different approaches for this: firstly, it applies a machine learning classifier and if it cannot recognize the license, it uses the detector based on the Sørensen-Dice coefficient.

\subsubsection{Machine learning classifier}
The first detector uses a machine learning classifier model. To train this model, we gathered a dataset of license texts from all Java projects in the Public Git Archive~\cite{PGA}. This dataset contains projects with more than 50 stars on GitHub, more details about the projects can be found in our previous work~\cite{JetBrains_Arcticle}. To label these files, we used \textsc{Scancode}~\cite{scancode_toolkit}. This tool is very accurate but rather slow, so we wanted to create a machine learning model that would learn the same labels and would be able to work faster. After the labelling was completed, we filtered out all licenses that had fewer than 100 files, in order to make the classifier more robust and focus on the most popular licenses. This resulted in 15,555 license files that altogether contain 12 most popular licenses. Their full list can be found in the README in the plugin's repository on GitHub~\cite{LDP_github}.

We trained the classifier in Python using the CatBoost library~\cite{catboost}. The text of each license was vectorized using \textit{CounterVectorizer}, we used a train-test split of 3:1, the resulting model reached the accuracy of 97\%. The model was transformed into the ONNX format~\cite{ONNX} and then converted into pure Kotlin for inference using the KInference~\cite{kinference} library. An analogous \textit{CounterVectorizer} was written in Kotlin, which allowed the plugin to infer the model in Kotlin without accessing other tools or languages, and thus work faster. 

\subsubsection{Sørensen-Dice coefficient}
If the ML detector could not recognize a given license, a second detector is used based on the Sørensen-Dice similarity coefficient. The similarity coefficient for two sets of elements $ X $ and $ Y $ is defined as: \[ DSC=\frac{2|X\cap Y|}{|X|+|Y|}, \] where $|X|$ and $|Y|$ are the corresponding cardinalities of sets. 

The detector based on the similarity coefficient works as follows. Each supported license contains a predefined set of words in the original text. The detector accepts the text of an unknown license as input and splits it into words. The resulting set of words is compared using the Sørensen-Dice coefficient with the word sets of the original licenses. The license counts as detected if the similarity coefficient is greater than the threshold value of 0.98. Thus, the detector requires 98\% coincidence with the original text of the license and allows for minor changes to the text. This threshold value was chosen, since it is used in \textsc{Licensee}~\cite{licensee}, another tool that employs the Sørensen-Dice coefficient for detecting licenses.

The detector based on the similarity coefficient supports the same 12 licenses that the ML detector supports, as well as 4 additional ones: CDDL-1.0, EPL-1.0, GPL-2.0-with-classpath-exception, and MPL-1.0. This detector is also very simple to extend: for a new license, it is only necessary to add its text and its name. The reason why \toolname currently does not support more licenses is that in order to detect incompatibilities, it is also necessary to specify compatible licenses for each new license. According to recent studies~\cite{JetBrains_Arcticle, vendome2015large}, 16 licenses that the plugin currently supports cover more than 95\% of projects and libraries with more that 50 stars on GitHub, which we consider to be acceptable.

\subsection{Detecting Licenses of Modules}

The extraction of the information about the licenses of the project's modules happens as follows. The IntelliJ Platform provides the information about all the modules in the project. \toolname scans the root directories of each module for a file that has a name commonly associated with a license file (\eg \textit{LICENSE.txt}). The discovered files contain the full text of the license, which is used to determine its type, as described in Section~\ref{sec:detecting_licenses}. This process results in the list of project modules with detected licenses. If a submodule has no license, then it is assumed to have the license of its parent module. For example, if the project only has the main license in the root module and no others, then this license is assumed to cover all of its submodules.

\subsection{Detecting Licenses of Libraries}
\label{sec:detecting_licenses_of_libraries}

The next task is determining the licenses of libraries. The libraries are usually stored in the form of \texttt{jar} archives. 
In this archive, there may be a license file (similar to the license of a module) with the full text of the library license. It can be used to determine the license type, as described in Section~\ref{sec:detecting_licenses}. The archive should also contain a  \textit{pom.xml} file, which contains meta-information about the library, including the name of its license. However, there are no strictly formalized names of licenses, so the name of the same license may differ in the metadata of different libraries. If the library archive contains this file, \toolname uses regular expressions to obtain a license name in the SDPX format.

Unfortunately, in practice, libraries are often supplied without the information about their license.
In this case, it is almost impossible to find out the license of the library without making requests to third-party resources.
To deal with this scenario, \toolname uses JetBrains Package Search.\footnote{JetBrains Package Search: \url{https://package-search.jetbrains.com/}}
This service indexes popular repositories and hostings for Java libraries and provides an API for obtaining complete information about a specific library, including the library license. 
A request to Package Search is made even when the license is already determined from the \texttt{jar} archive, because Package Search also provides a link to the library's homepage or GitHub repository that \toolname uses to provide the user with a possibility to learn more about the library.

\subsection{Finding Incompatibilities between Licenses}
\label{sec:incompatibilities}

When the licenses for all the modules and libraries are detected, \toolname can carry out its main task---detect licensing incompatibilities. The plugin can detect two types of incompatibilities:

\begin{enumerate}
    \item The license of the library is not compatible with the license of the module where this library is used.
    \item The license of a submodule is not compatible with the license of its parent module.
\end{enumerate}

They are detected as follows. Each possible license of a library has a predefined set of licenses that are compatible with it, \ie licenses that a module should have to use this library. Using these sets, each library in each module is checked whether the module's license is compatible with the library's license. If not, a potential violation is detected. For example, AGPL-3.0 is a strong copyleft license, meaning that libraries licensed under it can only be used in projects and modules with the same AGPL-3.0 license. Therefore, if the library with AGPL-3.0 is used in the module with a permissive BSD-3-Clause license, it is a violation.

The incompatibilities between modules and submodules are detected in a similar way.

\subsection{Suggesting the Appropriate License}
\label{sec:suggesting}

Sometimes developers are reluctant to add a license to their project, which is a problem in itself. Even permissive libraries like Apache-2.0, MIT, or BSD require the software that uses libraries covered by them to be open-source, so using them without any project license at all is almost always a violation. \toolname can help solve this problem by suggesting a proper license for the particular module or the whole project.

This is also done using the compatibility sets of licenses. The plugin gathers compatibility sets for all licenses of all libraries in the module and calculates their intersection. Since the licenses in the intersection are compatible with all the libraries, the plugin suggests them to the user. 

For example, if the project with no license is using two libraries, one of which has the MIT license and the other has the AGPL-3.0 license, the intersection of compatibility sets will include the AGPL-3.0 license, which will be suggested.

\section{User Interface}

\begin{figure}[t]
     \centering
     \includegraphics[width=\linewidth]{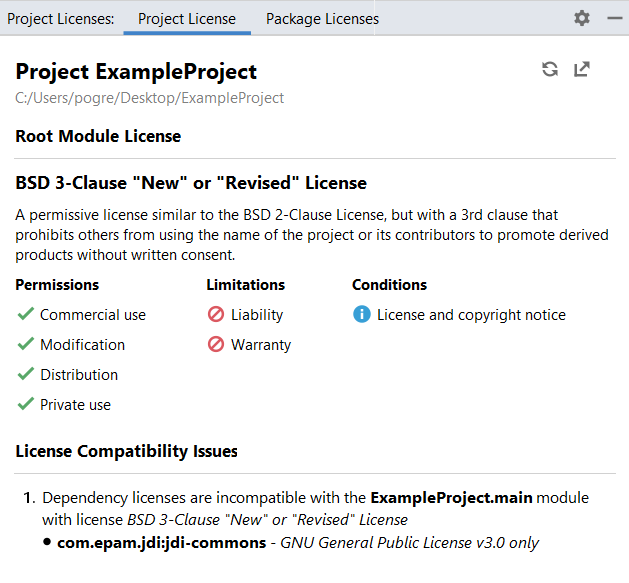}
     \caption{The \textit{Project License} tab in the tool window.}
     \label{fig:project}
\end{figure}

\toolname provides the user with a graphical interface for the convenience of managing licenses. 
The main graphical interface of the plugin is a \textit{Tool Window}. The window contains two tabs and provides the information about all the licenses in the project. The first tab is called \textit{Project License} and is presented in Figure~\ref{fig:project}. It contains the information about the detected main license of the project (root module), its description (permissions, limitations, and conditions), as well as a list of detected potential license violations. 

\begin{figure}[h!]
     \centering
     \includegraphics[width=\linewidth]{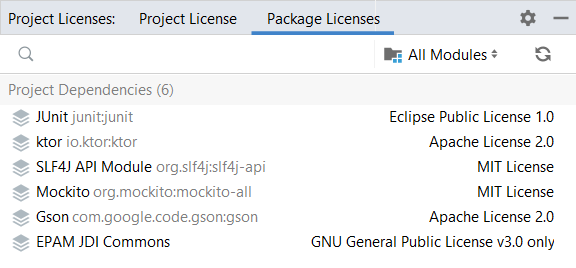}
     \caption{The \textit{Package Licenses} tab in the tool window.}
     \label{fig:libraries}
     \vspace{-0.3cm}
\end{figure}

The second tab is called \textit{Package Licenses} and is presented in Figure~\ref{fig:libraries}. This tab contains the information about all the licenses of libraries used inside the project, it supports a search among all the libraries and filtering by modules.

\begin{figure}[h!]
     \centering
     \includegraphics[width=\linewidth]{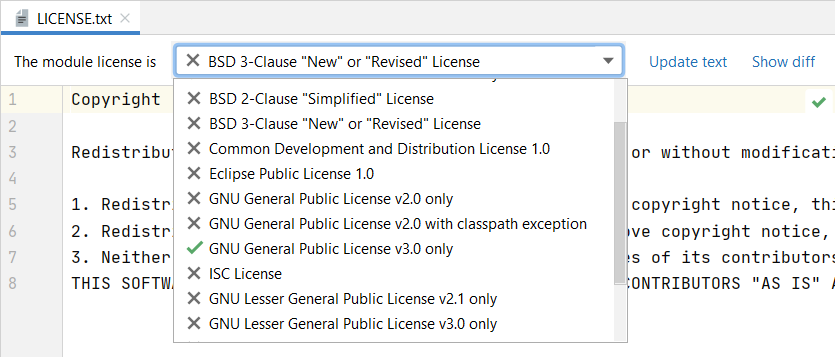}
     \caption{The \textit{License Editor Notification} panel. The green \checkmark sign indicates that this license is compatible with all the module's libraries.}
     \label{fig:editor}
     \vspace{-0.3cm}
\end{figure}

In addition to the Tool Window, the plugin provides the \textit{License Editor Notification} panel, presented in Figure~\ref{fig:editor}. This panel appears at the top of the editor when the module license is opened in it. The panel allows the user to change the license in several simple clicks using the drop-down menu, while indicating which licenses are compatible with all the licenses of the module's libraries. The panel also provides an opportunity to compare the current text of the license file with the original text of the license to check for possible differences.

Also, the plugin adds a new item into the IDE's \textit{New file...} menu, called \textit{Module License File}. If the project has no license, the user can use this functionality, and the plugin will detect the licenses compatible with all the project's libraries as described in Section~\ref{sec:suggesting}, and suggest the most permissive one to the user. This way, even the most inexperienced user can manage their licenses and not make mistakes. In the future, the user can always change their license using the \textit{License Editor Notification} panel described above.

Finally, the developed plugin provides another convenient way of viewing the licenses of libraries by providing hints right in the build system script. A hint with the name of the license of a given library appears next to each command that connects the library.
These hints allow the user to keep the licenses in mind when adding new libraries directly in the editor. The hints are implemented for Maven, Groovy Gradle, and Kotlin Gradle scripts.

\section{Evaluation}\label{sec:evaluation}

As a preliminary evaluation, we checked one of the the plugin's features---an ability to create an appropriate license file for a Java project without a license. Using the SEART~\cite{Dabic:msr2021data} tool, we have selected all Java projects on GitHub (excluding forks) that have between 10 and 100 stars. From a total of 62,247 such projects, 30,876 (49.6\%) projects did not have a license, which indicates the importance of the proposed functionality. We randomly selected 60 projects that have no license. We opened each project in IntelliJ IDEA and using the plugin created a license file that would satisfy the licenses of all libraries in the project. The plugin correctly created a file with a compatible license for 56 projects out of 60 (93\%).

In 4 cases, the plugin was unable to select a license that satisfied all the restrictions. 
The main reason for this is that the information about library licenses that the plugin receives from Package Search sometimes contains inaccuracies. In all 4 cases, the service provided an incorrect \textit{GPL-2.0} license to one of the project libraries instead of the true \textit{GPL-2.0-with-classpath-exception} license. Because of this, one library was licensed under GPL-2.0, and another---under GPL-3.0. In such cases, a contradiction arises and it is impossible to choose a suitable license for the project.

In the end, we believe that \toolname fulfills its task and can help developers create a correct license for a project in just several clicks with a decent accuracy, and therefore even an inexperienced developer will not make a mistake.

\section{Conclusion}\label{sec:conclusion}
In this paper, we present \toolname---a tool for managing licenses and detecting potential license incompatibilities. The tool was implemented as a plugin for IntelliJ IDEA and works with Java projects.
To determine a license type from its text, \toolname uses two detectors: a machine learning model and a detector based on the Sørensen-Dice similarity coefficient. Currently, the plugin supports 16 most popular Java licenses.
Using the information about licenses in the project, the plugin can detect potential incompatibilities between them and suggest possible license for the project if it has none.

\toolname can be further extended in a number of ways. We plan to add the support for more licenses (including the case of double licenses) and conduct a full-scale evaluation among users to see what kind of violations our tool is able and is not able to discover in the projects of small teams and individual users. It is also of interest to perfect the machine learning classifier in the tool: more experiments are needed to obtain the optimal thresholds, moreover, the detection thresholds can be changed by the user to indicate how conservative or how risky they want the detection to be. Finally, we plan to support \toolname's functionality in continuous integration as a part of JetBrains Qodana~\cite{qodana}.

\balance

\bibliographystyle{ieeetran}
\bibliography{paper}

\end{document}